\documentclass[a4paper,12pt]{article}
\usepackage{graphicx}
\usepackage[top=3.0cm,bottom=3.0cm,left=3.0cm,right=2.0cm]{geometry}
\usepackage{amssymb,amsmath}
\usepackage{textcomp}
\usepackage{subfigure}
\usepackage{cite}
\usepackage{braket}
\usepackage{xcolor}

\newcommand{\QCD}{{\textrm{\scriptsize QCD}}}

\title{Fractal structure in Yang-Mills fields and non extensivity}
\author{Airton Deppman$^{1,2}$, Eugenio Meg\'ias$^{2}$, D\'ebora P. Menezes$^{3}$}
\date{1- Instituto de F\'isica -  Universidade de S\~ao Paulo \\ email: deppman@if.usp.br ; \\  2- Departamento de F{\'\i}sica At\'omica, Molecular y Nuclear and
  Instituto Carlos I de F\'{\i}sica Te\'orica y Computacional - Universidad   de Granada; \\3- Departamento de F\'{\i}sica - Universidade Federal de Santa Catarina.}

\begin{document}

\maketitle

\begin{abstract}
    Scaling properties of Yang-Mills fields are used to show that fractal structures are expected to be present in system described by those theories. We show that the fractal structure leads to recurrence formulas that allow the determination of non perturbative effective coupling. Fractal structures also cause the emergence of non extensivity in the system, which can be described by Tsallis statistics. The entropic index present in this statistics is obtained in terms of the field theory parameters. We apply the theory for QCD, and obtain the entropic index value, which is in good agreement with values obtained from experimental data. The Haussdorf dimension is calculated in terms of the entropic index, and the result for hadronic systems is in good agreement with the fractal dimension accessed by intermittency analysis of high energy collision data. The fractal dimension allow us to calculate the behavior of the particle multiplicity with the collision energy, showing again good agreement with data.
\end{abstract}

\section{Main}
Fractals are complex systems with internal fine structure featuring scale invariance and self-similarity. Fractal measures, contrary to more conventional quantities for which an increase in resolution results in the same measured value with increased precision, yield different values for different resolutions. A classical example is the lengths of coastlines~\cite{Mandelbrot},  which increase significantly if measured with smaller scales: if measured in centimeters, the value is quite different from the one obtained if measure in kilometers.
%Although being a complex structure, fractals are usually obtained from iteration over a simple rule or pattern.
The concept of fractal has found applications in Mathematics, Biology and Physics, among other areas. It has been associated to socioeconomics evolution of cities, as well as to the actual configuration of urban area infrastructure~\cite{West}. It can emerge from the system structure or from its dynamical evolution. In geometry, it introduces a new kind of topology and renders beautiful pictures. An interesting introduction to the subject can be found in Ref.~\cite{Falconer}. 

Scale invariance is a fundamental aspect of the Yang-Mills field (YMF) theory, playing an important role in the renormalization of the theory after divergences are cut~\cite{Dyson1, Ward, Gell-Mann_Low}. This is a prototype theory for the description of three among the four known interactions: electromagnetic, weak and strong. The Standard Model is the most complete implementation of the theory  so far, encompassing strong, weak and electromagnetic forces
 found in Ref.~\cite{YMF-60y}. In the case of Quantum Chromodynamics (QCD),  used to describe 
the strong interaction, the scale invariance is expressed through the Callan-Symanzik equation~\cite{Callan, Symanzik1, Symanzik2}, which is a basic tool to show that QCD is an asymptotic free interaction~\cite{Politzer1, Politzer2, GrossWilczek1, GrossWilczek2}. This aspect of QCD is of fundamental importance in particle physics, since it means that for processes with high momentum transfer, as in deep inelastic scattering~\cite{DIS}, quarks are almost free and the scattering can be studied in low-order perturbative calculation. Thus, scale invariance means that the coupling constant, which represents the strength of the interaction, depends on the energy scale for the interacting particles. Scale dependent couplings are called effective couplings or running couplings.

In this work we show that fractal structures can be formed by systems described by the Yang-Mills field theory. The presence of fractal structures lead to a recurrence formula that allows the determination of the effective coupling even in high perturbative orders. A consequence of the fractal structure is that the proper thermodynamical theory for describing the interacting system is the non extensive Tsallis statistics~\cite{Tsallis1988, TsallisBook}, rather than the traditional Boltzmann-Gibbs statistics, where the entropic index, q, is a measure of the non additivity of the entropy. Here  $q$ is obtained, for the first time, in terms of the field theory parameters. The fractal dimension is determined as a function of the entropic index. When applied to QCD in the asymptotic approximation, the theoretical valued obtained for $q$ is in good agreement with the values found in the analyses of experimental data. We show that various experimental features
can be explained by the theoretical results derived here, in particular the behavior of particle multiplicity as a function of the collision energy, which depends on the fractal dimension.

Although fractal structure 
could be perceived in any field theory of Yang-Mills type, in QCD its effects are more evident. No wonder that hints of fractal structure were present in theories proposed already in the 1960's. About fifty years ago, Rolf Hagedorn proposed a thermodynamical description of high energy collisions based on a self-consistent principle~\cite{Hagedorn0,Hagedorn65} that states that fireballs, the transient hot system formed in those collisions, is made of ... fireballs! Although the definition of fireballs is based on the concept of fireball itself, Hagedorn was able to obtain a complete thermodynamic physical description for this system. The resemblance between the definition of fireball and that of fractals is clear, and we suppose that, were the concept of fractals already known at that time, Hagedorn would have used it for his definition of fireball. A little later, Chew and Frautschi proposed the bootstrap model for hadrons~\cite{Frautschi, Chew}, where it is supposed that hadrons are made of hadrons, showing again the similitude to fractals. Among the predictions of these theories are the existence of a limiting temperature, known as Hagedorn temperature; a formula for the mass spectrum of hadrons 
; an exponential distribution of energy and momentum of the particles.
These phenomenological theories obtained great success in the first years after their proposals. An entire new line of research emerged from these results, and many Hadron Resonance Gas Models are still used today~\cite{Petreczky_HRG, Megias_HRG, Venugopalan_HRG}. Despite the initial success, after the energy available for collisions in accelerators increased, it was soon realized that the exponential distribution predicted by the theory was in disagreement with experiments. With the success of QCD, those phenomenological approaches were, to some extent, dismissed. As is shown in the present work, our results reconcile Hagedorn theory with QCD.

Our starting point is the scaling property of YMF, which is expressed in terms of amplitudes or vertex functions, $\Gamma(p,m,g)$, as
\begin{equation}
  \Gamma(p,m,g)=\lambda^{-D} \Gamma(p,\bar{m},\bar{g}) \,,
\end{equation}
where $p,m,g$ are, respectively, the momentum, mass and coupling constant associated to a state in a non interacting configuration, as for instance, for systems that are so apart of each other that interaction is null in all practical aspects. The effective mass and effective coupling, $\bar{m}$ and $\bar{g}$ respectively, are obtained when interaction is considered. A direct consequence of the equation above is the renormalization group equation
\begin{equation}
  \left[M\frac{\partial}{\partial M}  + \beta_g \frac{\partial}{\partial \bar{g}} + \gamma \right]\Gamma=0 \label{CallanSymanzik} \,,
\end{equation}
where 
\begin{equation}
    \beta_g=M\frac{\partial \bar{g}}{\partial M}
\end{equation}
is known as beta function and gives the logarithmic ratio in which the coupling varies with the scale, and $\gamma$ is related to the change of scale of the interacting fields.

Consider an initial free state $\ket{\Psi_o}$. Its time dependent evolution is given by
\begin{equation}
 \ket{\Psi}= e^{-iHt} \ket{\Psi_o}\, \label{Gamma}
\end{equation}
where $H$ is the Hamiltonian operator. In the irreducible representation, each interaction is represented graphically by a proper vertex, while lines represent effective partons for which self-interaction is included through an effective mass. In the following, parton means effective parton, and interaction means proper vertex.

We introduce new states, $\ket{\Psi_n}$, such that
\begin{equation}
 \ket{\Psi_n} = (-i)^n \int dt_n \dots dt_1 g e^{-iH_o(t_n-t_{n-1})} g e^{-iH_o(t_{n-1}-t_{n-2})} g \dots e^{-iH_o(t_1-t_o)}  \ket{\Psi_o}
\end{equation}
and
\begin{equation}
 \ket{\Psi}=\sum_{\{n\}} \braket{\Psi_n|\Psi} \ket{\Psi_n}\,.
\end{equation}
These new states are those with a well defined number of vertexes, or interactions. The sum in equation above is taken over all possible configurations with $n$ vertexes.
The number of particles created is not necessarily equal to $n$, since some of the particles are created at high orders of perturbative calculation. So we introduce the states $\ket{\psi_N}$ with well defined number of particles created, $N$. These states are 
\begin{equation}
 \ket{\Psi_n}=\sum_{N} \braket{\psi_N|\Psi_n} \ket{\psi_N}\,. 
\end{equation}
The states $\ket{\psi_N}$ are autovectors of $H_o$ with fixed number of particles, $N$. Of course $\braket{\Psi_n|\psi_N}=0$ whenever $N > M(n)$, and $\braket{\Psi_{N'}|\Psi_N}=\delta_{N'N}$.
Since the number of partons is fixed and they do not interact  but by contact interaction, the states $\ket{\psi_N}$ can be understood as the states of an ideal gas of $N$ partons. Therefore
\begin{equation}
 \ket{\psi_N}={\cal S} \ket{\gamma_1,m_1,p_1, \dots , \gamma_N,m_N,p_N}\,,
\end{equation}
where $m_i$ and $p_i$ are the mass and momentum of the $i$ partonic state, and $\gamma_i$ represents all relevant quantum numbers necessary to completely characterize the partonic state. ${\cal S}$~is the symmetrization operator acting over fermions and bosons. In the case of hot systems the momentum of the effective partons varies continuously, ${\cal S}$ gives a negligible modification of the single parton states, so mass and momentum of each parton can vary independently, as far as the total energy is conserved.

As a consequence of the large number of possible ways to obtain $N$ particles in the final state, the actual process that leads to a particular state $\ket{\psi_N}$ is not relevant, and statistical methods can be applied. For instance, the probability to find a state where at least one parton has mass between $m_o$ and $m_o+dm_o$, and momentum coordinates  between $p_{oi}$ and $p_{oi}+dp_{oi}$, is given by matrix elements of the type
\begin{equation}
    P(\varepsilon_o)=\braket{\gamma_o,m_o,p_o, \dots|\Psi(t)}\,,
\end{equation}
where $\varepsilon_o=p_o^0$ is the energy of one parton in a system with an arbitrary number of partons. Initially, we assume the particles form an ideal gas and ignore that they have a fractal structure. We also assume, for reasons that will become clear below, that the energy of the $N$ particle system may fluctuate according to a probability density $P(E)$. In this case we can show that
\begin{equation}
 %\braket{\gamma_o,m_o,p_o \dots|\Psi}
 P(\varepsilon_o)= \sum_n  \sum_{N} G^n
  \left(\frac{N}{n(\tilde{N}-1)}\right)^4 \left(1+\frac{\varepsilon_o }{E}\right)^{-(4N-5)} P(E) \,, % d^4\left(\frac{p }{E}\right) dE\,.
  \label{probdensideal}
\end{equation}

\begin{figure}[t!]
 \centering
   \begin{subfigure}[]
        \centering
        \includegraphics[scale=0.3]{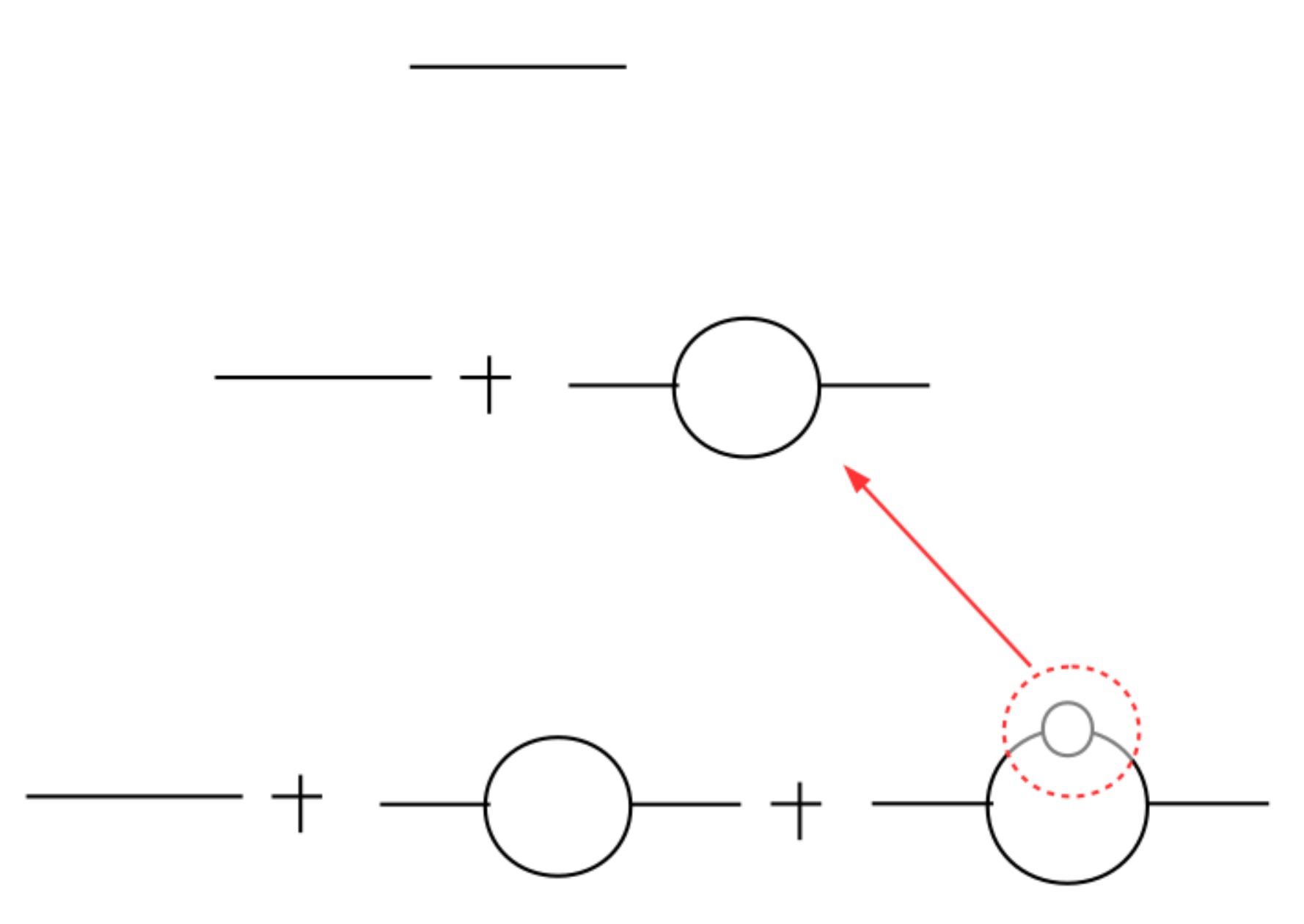}
   \end{subfigure}
   \begin{subfigure}[]
        \centering
        \includegraphics[scale=0.45]{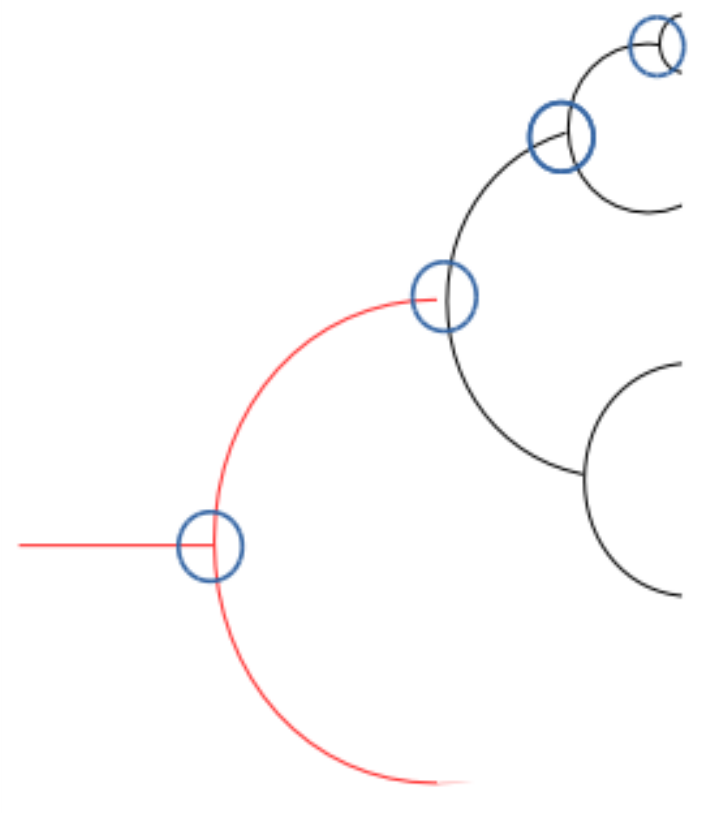}
   \end{subfigure} 
        \caption{(a) Scaling property of Yang-Mills fields as observed in diagrammatic representation. Loops and partons at different perturbative orders are equal, provided the appropriate scales are considered. (b) Schematic view of the fractal structure. The initial parton (first black line form the left) may be considered as a constituent of another parton (first red line from the left). At the vertexes highlighted by blue circles, the distribution of energy and momentum of the generated partons are determined. The region in black lines was considered in the text as an ideal gas with $N$ particles.}\label{fractalstructure}
\end{figure}

The assumption that the particles have no internal structure and that the system behaves like an ideal gas is not strictly valid, since in the actual case  particles present a fine structure with scaling and self-similarity, that is, they have a {\it fractal structure} (see Figure~\ref{fractalstructure}). The internal degrees of freedom share part of the total energy available, and thus distort the distribution of energy. Our next step is to include the effects of the fractal structure into the distribution obtained above.
The probability density $P_N$, when it is written in terms of scale independent quantities, must be identical for all effective partons, since these systems are self-similar objects, therefore $P_N$ must be independent of the number of particles, $N$, and on the order of perturbative calculation, $n$. The scaling property also imposes that the ratio between the energy $\varepsilon_o$ of a parton and that of its parent parton, $E$, must be the same for all partons, independently of $N$ and~$n$. Therefore we can write \begin{equation}
 \frac{\varepsilon_o}{E} \sim \frac{E}{{\cal M}}\equiv\frac{\varepsilon}{\Lambda} \,, 
\end{equation} 
with $\varepsilon/\Lambda$ being independent of the level in the fractal structure. The self-similarity among the partons implies that the probability that the parent parton with mass $E$ inside a larger system with mass ${\cal M} $ is similar to the probability given in Eq.~(\ref{probdensideal}), as represented in Figure~\ref{fractalstructure}(a), so we write 
\begin{equation}
  P\left(E\right)=\left(1+\frac{E}{{\cal M}}\right)^{-\alpha} \,, \label{power-law}
\end{equation} 
where, $\alpha$ is the number of degrees of freedom of the fractal structure in the system with energy ${\cal M}$. Since part of the fractal structure corresponds to the $N$ particles we are considering apart, only a fraction $\nu$ of the degrees of freedom that remains to be added to the fractal structure
, as schematically shown in Fig~\ref{fractalstructure}(b). Then we can write for the probability density
\begin{equation}
 %\braket{\gamma_o,m_o,p_o \dots|\Psi}
 P(\varepsilon_o)= \sum_n  \sum_{N} G^n
  \left(\frac{N}{n(\tilde{N}-1)}\right)^4 \left(1+\frac{\varepsilon_o }{E}\right)^{-(4N-5)} \left[P(E)\right]^{\nu} \,, % d^4\left(\frac{p }{E}\right) dE\,.
  \label{probdensfractal}
\end{equation}
which includes some characteristics of the ideal gas behavior, but takes into account fractal degrees of freedom.

Substituting Eq~(\ref{power-law}) into Eq.~(\ref{probdensideal}), and using self-similarity of fractals to identify $P(\varepsilon_o)=P(E)$, it results
\begin{equation} 
(4N-5)+\alpha \nu=\alpha \,, 
\end{equation} 
where the parameter $\nu \leq 1$ represents the fraction of the total number of degrees of freedom of the state $\ket{\psi_N}$, when it is considered as a constituent of a larger system. We can introduce the parameters $q$ such that 
\begin{equation} 
 \frac{1}{q-1}=\frac{4N-5}{1-\nu}\,, \label{qvalue} 
\end{equation} 
with $q>1$, and $\Lambda$ such that 
\begin{equation} \Lambda=\frac{4N-5}{1-\nu} \lambda\,, \label{resolution} \end{equation} 
where $\lambda$ is a reduced scale, independent of the number of degrees of freedom relevant to the system, which gives the energy per degree of freedom and therefore can be interpreted as a kind of temperature. In terms of the new parameters we obtain 
\begin{equation}
  P(\varepsilon/\lambda)=\left[1+(q-1)\frac{\varepsilon}{\lambda}\right]^{-\frac{1}{q-1}}\,. \label{Tdistribution} 
\end{equation} 

This result shows that the distribution of parton energy  in a system governed by Yang-Mills fields depends only on the ratio between the parton energy, $\varepsilon$, and the energy scale per degree of freedom $\lambda$. Furthermore, the right hand side of Eq.~(\ref{Tdistribution}) is exactly the q-exponential distribution commonly found in Tsallis non extensive statistics. The non additive entropy as the basis for a non extensive statistics was proposed by Tsallis in the late 80's \cite{Tsallis1988,TsallisBook} and its effects have been explored since then, but they are not completely understood yet. 

The $q$-deformed entropy functional that underlines non extensive statistics depends on a real parameter, $q$, that determines the degree of nonadditivity of the functional, and in the limit $q \rightarrow 1$ it becomes additive and the standard Boltzmann-Gibbs entropy is recovered. Similar results have been obtained through a different approach using the concept of thermofractals, introduced in Ref.~\cite{Deppman2016} and studied in details in Ref.~\cite{DFMM}. There it is shown that the fractal structure leads to the non extensive statistics, and it is discussed the relations between thermofractals and Hagedorn's self-consistent thermodynamics developed to study high energy collisions~\cite{Hagedorn0, Hagedorn65}, and that was extended to non extensive statistics~\cite{Deppman2012}.  Thus the fact that the non extensive statistics is obtained from considerations of the scaling properties and self-similarity in Yang-Mills fields, shows that Tsallis statistics should have been used in Hagedorn's theory, instead of Boltzmann-Gibbs statistics, and this is the reason Hagedorn's theory fails when confronted with high energy colisions data. Indeed, when the self-consistent thermodynamics is obtained by using the non extensive statistics it can describe more accurately the experimental observations of hadronic systems~\cite{Deppman2012}.  Our results show that Tsallis statistics emerges in YMF from the fact that the fields can be quantized into partons only if these partons present a complex internal structure similar to fractals. The meaning of the entropic index, in this case, is clear: Eq~(\ref{qvalue}) shows that $(q-1)^{-1}$ gives the number of degrees of freedom relevant in the system. Observe that as the number of degrees of freedom increases, $q \rightarrow 1$ and Boltzmann statistics is recovered as a limiting case. However, we will see that $q$ can be calculated from the field theory parameters, and that for QCD the number of degrees of freedom is finite, so Tsallis statistics must be used.

The role played by the q-exponential in Eq.~(\ref{Tdistribution}) is that of determining, stochastically, the values of effective masses and momenta of the particles generated at each vertex (see Fig~\ref{fractalstructure}b). Therefore, the effective vertex should contain a term which is given by that q-exponential function, that is
\begin{equation}
  g=\prod_{i=1}^{\tilde{N}}G\left[1+(q-1)\frac{\varepsilon_o}{\lambda}\right]^{-1/(q-1)}\,, \label{coupling}
\end{equation}
we can compare the behavior of the effective coupling given above with the expected behavior of the QCD coupling as calculated in 1-loop approximation, where perturbative methods can be used to obtain the $\beta$-function. From Eq.~(\ref{coupling}) we get 
\begin{equation} 
  \beta_{\bar{g}}=-\frac{1}{16\pi^2} \frac{1}{q-1} g^{\tilde{N}+1}\,, \label{betaFractal} 
\end{equation} 
and we emphasize that $\tilde{N}=2$. The beta-function for QCD is~\cite{Politzer1} 
\begin{equation} 
  \beta_{\QCD}=- \frac{g^3}{16\pi^2} \left[\frac{11}{3}c_1-\frac{4}{3}c_2\right]\,, \label{betaQCD} 
\end{equation} 
where the parameters $c_1$ and $c_2$ are related to the number of colors and flavors by $c_1=N_c$ and $c_2=N_f/2$. Using $N_c=N_f/2=3$ it results 
\begin{equation}
 \frac{11}{3}c_1-\frac{4}{3}c_2=7\,, 
\end{equation} 
which, by comparison of Eqs.~(\ref{betaFractal}) and~(\ref{betaQCD}), leads to $q=1.14$. From experimental data analysis\cite{Cleymans,WilkWlodarczyk,Lucas1, Lucas2, Sena} $q=1.14 \pm 0.01$, showing  good agreement between theory and experiments. The agreement found between experimental and theoretical values for $q$ shows that the effective coupling obtained from considerations on the fractal structure of QCD is  good agreement with the accumulated knowledge about strong interaction. The advantage of the effective coupling described by Eq.~(\ref{coupling}) is that it may be used in non perturbative calculations. Furthermore, it represents a reconciliation between QCD and Hagedorn's self-consistent approach.

The results obtained here have shown that a system with fractal structure, similar to the thermofractals~\cite{Deppman2016}, can be understood as a natural consequence of the scale invariance of gauge field theories. We give solid grounds for phenomenological approaches that have been used to describe hadron mass spectrum~\cite{Lucas1} and  multiparticle production with non extensive statistics~\cite{Cleymans, WW2018, Lucas1, Lucas2, Sena, WilkWlodarczyk, DeBhaskar, Parvan, Ishihara, Tawfik, Rybczynski}, which can explain the long tail distribution observed in multiparticle production. From the recurrence method used here we understand the reason why, even at small order of perturbative QCD calculation, it is possible to describe correctly the transversal momentum distributions measured at high energies by adopting Tsallis distribution~\cite{Wong}. The fact that an statistical interpretation of the field theory is possible, allows us to understand why the non extensive self-consistent theory is in good agreement with Lattice QCD calculations~\cite{Deppman2014}.

It is possible to understand, from the considerations made here, that the fractal structure of YMF is the basis for investigations of hadron properties~\cite{PedroCardoso}, phase-transition in hot hadronic matter~\cite{Megias}, neutron stars~\cite{Debora} and cosmic-ray~\cite{BeckCosmicRay}.  These phenomenological approaches are, in fact, implementations of  scaling symmetries observed in Yang-Mills fields. The fractal structure also allows the understanding of the self-similarity~\cite{WWselfsymmetry, Tokarev, Zborovsky} and scaling properties observed in high energy experimental data. In fact, these findings are direct consequences of the scaling properties of YMF, as discussed here.
Moreover, the fact that the entropic index, $q$, is obtained from well-known field-theoretical parameters, the results we have obtained allows a new interpretation of Tsallis statistics in terms of fractal structure in the same lines it was obtained in thermofractals approach~\cite{Deppman_Universe}.

The fractal structure presents at least one fractal dimension, the Haussdorf dimension, which was calculated for the case of thermofractals in Ref.~\cite{Deppman2016}.
The Haussdorf dimension can be calculated by using the box-counting technique~\cite{Falconer}, where the dimension $D$ is related to the number of boxes, ${\cal N}$, necessary to completely cover all possible values for the measured quantity, whatever it is, and $D_t$ is the topological dimension, that is, the dimension expected for some quantity when there is no fractal dimension. At some scale $r$ these quantities are related by\cite{Falconer}
\begin{equation}
    {\cal N} r^{-D} \propto r^{-D_t}\,.
\end{equation}
In our case, $D_t=1$ is the dimension that describes how the total energy varies when the energy unit, $r$ is modified, and ${\cal N}=\tilde{N}^n$, where $n$ is the layer in the fractal structure where the partons have energies of the order of $\lambda$. $D$ is how the energy of the system components varies with $r$. We determine $D$ by
noticing that the average parton energy at a scale $\lambda$ is 
\begin{equation}    
\braket{\varepsilon}=\frac{\lambda}{2q-1}\,. 
\end{equation} 
Then, it follows that the fractal dimension is 
\begin{equation} 
  D-1=\frac{\log \tilde{N}}{\log R}\,, 
\end{equation} 
where $R=(q-1)/(2q-1)$ is the ratio between the average energy of the constituent fractal and the energy of its parent system. Using $q=1.14$ and $\tilde{N}=2$, it results $D=0.69$, in good agreement with findings from intermittency analysis of high energy collision data~\cite{Bialas_Peschanski, Bialas_Peschanski2, Hwa, HwaPan, Hegyi1,DreminHwa,Hegyi2,Antoniou}. The fractal dimension gives the behavior of the parton energy with the energy scale, $r$, that is, while the total energy goes as $E \propto r^{-1}$, the partons observed at scale $\lambda$ have energies that depend on the scale as $\varepsilon \propto r^{-D}$. A more direct way to access the fractal dimension is the particle multiplicity. In fact, being $\cal{M}$ the particle multiplicity, we have 
\begin{equation} 
  {\cal M} \braket{\varepsilon}=E \,. 
\end{equation} 
From the dimensional behavior obtained above, we get 
\begin{equation} 
  {\cal M}=E^{1-D}\,. 
\end{equation} 
For the case of hadrons, as we have seen, $q=1.14$ and $D=0.69$, so we obtain ${\cal M}\propto E^{0.31}$, which is in excellent agreement with the result obtained for $pp$ collision at high energy~\cite{Sarkisyan_Multiplicity}, which gives, for a power-law fit, an exponent corresponding to $1-D=0.302$. 

In conclusion, we used scaling properties of Yang-Mills fields to show that fractal structures are expected to be formed in systems described by that theory. These structures lead to a thermodynamic description of the fields that follows Tsallis statistics, with the entropic index, $q$, being for the first completely determined in terms of the field theory parameters. We discuss that Hagedorn's self-consistent theory fails because the non extensive effects were not considered, and verify that our results confirm the phenomenological extension of Hagedorn theory by adopting Tsallis statistics, which was proposed in previous works in the context of thermofractals.

Another consequence of the fractal structure is that we obtain a recurrence formula which allows the determination of the effective coupling in terms of particle momentum and of a scale. The fractal dimension is also obtained in terms of $q$, and therefore also in terms of the field theory parameters. We use QCD as an exemple of application of the theory. The beta function describing the behavior of the effective coupling can be obtained for QCD in 1-loop approximation, due to asymptotic freedom, and in such approximation we compare the results of our theory with the begavior expected for perturbative QCD, obtaining a good agreement. As a result, we obtain $q=1.14$, which is in good agreement with the value obtained from the analysis of experimental data from high energy collisions.

Haussdorff fractal dimension is calculated in terms of $q$, and for QCD results in $D=0.69$, which is a value in agreement with those obtained from intermittency analyses for $pp$ collisions. The fractal dimension is a manufestation of non extensivity which appears clearly in the behavior of the particle multiplicity as a function of the collisions energy. We show that multiplicity increases as a power-law of the collision enegly with exponent $1-D$, which gives 0.31 for QCD, again in good agreement with experimental data.

With the theory developed here, we explain self-similarity effects observed in multi-particle production in high energy collisions, as well as scaling properties of distributions obtained experimentally. Our results give a strong basis for phenomenological studies that have been used so far to explain experimental data from high energy collisions, as well as for phenomenological models that have been used to describe hadronic systems, such as nucleon structure and neutron stars. It also allows for a better understanding of the origins of non extensivity in high energy collisions and on the role of the entropic parameter, which is related here to the number of relevant degrees of freedom of the fractal structure.

\section{Method}
Consider that at an initial instant $t_0$  we have the system in an initial state $\ket{\Psi_o}$ which can be considered {\it free} in the sense that no interaction can take place. The time-dependent states $\Psi$ can be written as 
\begin{equation} 
  \ket{\Psi}=e^{-iHt}\ket{\Psi_o}\,, 
\end{equation} 
where $H$ is the Hamiltonian operator. If $H_o$ is the Hamiltonian for free effective partons, for which self-interaction is already included by considering that the parton has an effective mass and interacts according to an effective coupling~\cite{Dyson1, Gell-Mann_Low}. The state $\ket{\Psi}$ can be written as
\begin{equation} \ket{\Psi}=\sum_{\{n\}} \braket{\Psi_n|\Psi} \ket{\Psi_n}\,, \end{equation} 
where $\sum_{\{n\}}$ represents the sum over all possible irreducible graphs with $n$ interactions with the states $\ket{\Psi_n}$ corresponding to a fixed number, $n$, of interactions in the vertex function, such that 
\begin{equation}
 \ket{\Psi_n} = (-i)^n \int dt_n \dots dt_1 g e^{-iH_o(t_n-t_{n-1})} g e^{-iH_o(t_{n-1}-t_{n-2})}  \dots g e^{-iH_o(t_1-t_o)}  \ket{\Psi_o}
\end{equation}
with $t_n>t_{n-1}>\dots>t_1$. Of course these states satisfy the relation $\braket{\Psi_{n'}|\Psi_n}=\delta_{n'n}$.

The number of particles in the state $\ket{\Psi_n}$ is not directly related to $n$, since high order contributions to the $N$ particles states can be important, but certainly $N \le M(n):=n(\tilde{N}-1)+1$, where $\tilde{N}$ is the number of particles created or annihilated at each interaction. In Yang-Mills field theory $\tilde{N}=2$.\footnote{Diagrams with four external lines represent contact interaction and are not considered, since they give a non renormalizable contribution. However, when all diagrams are summed up, the contribution of the contact interaction is null.} We can introduce states of well defined number of effective partons, $\ket{\psi_N}$, so that 
\begin{equation} \ket{\Psi_n}=\sum_{N} \braket{\psi_N|\Psi_n} \ket{\psi_N}\,. \end{equation} 
The states $\ket{\psi_N}$ are autovectors of $H_o$ with fixed number of particles, $N$. Of course $\braket{\Psi_n|\psi_N}=0$ whenever $N > M(n)$, and $\braket{\Psi_{N'}|\Psi_N}=\delta_{N'N}$. Since the number of partons is fixed and they do not interact but by contact interaction, the states $\ket{\psi_N}$ can be understood as the states of an ideal gas of $N$ partons. Therefore 
\begin{equation} 
  \ket{\psi_N}={\cal S} \ket{\gamma_1,m_1,p_1, \dots , \gamma_N,m_N,p_N}\,, 
\end{equation} 
where $m_i$ and $p_i$ are the mass and momentum of the $i$ partonic state, and $\gamma_i$ represents all relevant quantum numbers necessary to completely characterize the partonic state. ${\cal S}$~is the symmetrization operator acting over fermions and bosons. Since the mass of the effective partons varies continuously, ${\cal S}$ gives a negligible modification of the single parton states, so mass and momentum of each parton can vary independently, as far as the total energy is conserved. 

Assuming that a statistical approach can be used, the probability to find a particle with quantum numbers, mass and momentum given, respectivelly, by $\gamma_o$, $m_o$ and $p_o$ is
\begin{equation} 
 \braket{\gamma_o,m_o,p_o, \dots|\Psi(t)}=\sum_n \sum_{N} \braket{\Psi_n|\Psi(t)} \braket{\psi_N|\Psi_n} \braket{\gamma_o,m_o,p_o,\dots|\psi_N}\,. \label{1partonprob1}
\end{equation}

The first bracket depends on the intensity of the interaction, determined by a coupling constant $G$, so 
\begin{equation}
  \braket{\Psi_n|\Psi}=G^n P(E) dE\,, \label{braket1}
\end{equation} 
where $P(E) dE$ is the probability to have the $N$ particle system with energy between $E$ and $E+dE$. The second bracket depends on the relative number of possibilities to get the configuration with $N$ particles, so 
\begin{equation} 
 \braket{\psi_N|\Psi_n} = C_N(n)\, \label{braket2}
\end{equation} 
with 
\begin{equation} \sum_{n} C_N(n)=1 \,, 
\end{equation} 
with $n \ge N/(\tilde{N}-1)$.
The last bracket in the expression above is calculated statistically by supposing that all possible configuration of the system have the same probability, and by counting the number of configurations with energy between $E$ and $E+dE$ and one particle with quantum numbers $\gamma_o$, effective mass $m_o$ and momentum $p_o$. . The result is 
\begin{equation} f(p_o) d^4p_o = A(N) P_N\left( \frac{\varepsilon_o}{E} \right) d^4\left( \frac{p_o}{M} \right) \,, 
\end{equation} 
where 
\begin{equation} P_N\left( \frac{\varepsilon_o}{E} \right) = \left( 1 - \frac{\varepsilon_o}{E} \right)^{4N-5}\,,  \label{braket3}
\end{equation} 
with 
\begin{equation} 
  A(N)=\frac{\Gamma(4N)}{8\pi\Gamma(4(N-1))}\,, \label{normalconst}
\end{equation} 
and $\varepsilon_o=p_o^0$ being the the particle energy. In most cases, including the cases of interest here, $\varepsilon \ll E$. Noticing that
\begin{equation}
    \left(1-\frac{\varepsilon_o }{E}\right)^{4N-5}= \left(1+\frac{\varepsilon_o }{E}\right)^{-(4N-5)} \left(1-\frac{\varepsilon_o^2 }{E^2}\right)^{4N-5}\,,
\end{equation}
and observing that the last term in the right-hand side is approximately the unit, we can write
\begin{equation}
    \left(1-\frac{\varepsilon_o }{E}\right)^{4N-5}= \left(1+\frac{\varepsilon_o }{E}\right)^{-(4N-5)}\,. \label{chgsignals}
\end{equation}

Observe that $A(N) \propto (4N)^4$ for $N$ sufficiently larger than unit, so the number of possible configurations increase fast with $N$. The maximum number of particle for a given number of interactions, $n$, is $M(n)=n(\tilde{N}-1)$, so we can consider the probability for  $N$ particles generated in $n$ interactions as
\begin{equation}
    C_N(n) \propto \left(\frac{N}{n(\tilde{N}-1)}\right)^4\,. \label{Ndependence}
\end{equation}

Substituting Eqs.~(\ref{braket1}),~(\ref{braket2}),~(\ref{braket3}),~(\ref{chgsignals}) and~(\ref{Ndependence})  into Eq.~(\ref{1partonprob1}), we obtain Eq.~(\ref{probdensideal}).
This result shows that for an ideal gas with finite number of particles, the probability depends on a power-law function of the ratio $\varepsilon_j/E$. 

\vspace{1cm}

{\bf Acknowledgments:} A.D. would like to thank the University of Granada, where part of this
work has been done, for the hospitality and financial support under a
grant of the Visiting Scholars Program of the Plan Propio de
Investigaci\'on of the University of Granada. He also acknowledges the
hospitality at Carmen de la Victoria. A.D. and D.P.M. are
partially supported by the Conselho Nacional de Desenvolvimento
Científico e Tecnológico (CNPq-Brazil) and by Project INCT-FNA
Proc. No. 464898/2014-5. The work of E.M. is supported by the Spanish MINEICO
and European FEDER funds under Grants FIS2014-59386-P and
FIS2017-85053-C2-1-P, by the Junta de Andaluc\'{\i}a under Grant
FQM-225, and by the Consejer\'{\i}a de Conocimiento, Investigaci\'on y
Universidad of the Junta de Andaluc\'{\i}a and European Regional
Development Fund (ERDF) Grant SOMM17/6105/UGR. The research of E.M. is
also supported by the Ram\'on y Cajal Program of the Spanish MINEICO
under Grant RYC-2016-20678. The authors thank Dr. Tobias Frederico for reading the manuscript and for his suggestions and criticisms. A.D. is thankful to Dr. Jos\'e Ademir de Lima for discussions about the statistics of small systems.

\end{document}